# Unidirectional ultracompact DNA-templated optical antennas

*Fangjia Zhu[1][†], María Sanz-Paz[1][†], Antonio Fernández-Domínguez[2], Xiaolu Zhuo[3,4], Luis M. Liz-Marzán[3,4,5], Fernando D. Stefani[6,7]\*, Mauricio Pilo-Pais[1]\*, and Guillermo P. Acuna[1]\**

[1] Department of Physics, University of Fribourg, Chemin du Musée 3, Fribourg CH-1700, Switzerland.

[2] Departamento de Física Teórica de la Materia Condensada and Condensed Matter Physics Center (IFIMAC), Universidad Autónoma de Madrid, E-28049 Madrid, Spain.

[3] CIC biomaGUNE, Basque Research and Technology Alliance (BRTA), Paseo de Miramón 182, Donostia-San Sebastian 20014, Spain.

[4] Centro de Investigación Biomédica en Red de Bioingeniería, Biomateriales y Nanomedicina (CIBER-BBN), Donostia-San Sebastian 20014, Spain.

[5] Ikerbasque, Basque Foundation for Science, 48009 Bilbao, Spain

[6] Centro de Investigaciones en Bionanociencias (CIBION), Consejo Nacional de Investigaciones Científicas y Técnicas (CONICET), Godoy Cruz 2390, C1425FQD Ciudad Autónoma de Buenos Aires, Argentina.

[7] Departamento de Física, Facultad de Ciencias Exactas y Naturales, Universidad de Buenos Aires, Güiraldes 2620, C1428EHA Ciudad Autónoma de Buenos Aires, Argentina

[†] both authors contributed equally

**Corresponding Authors**

\* E-mail: fernando.stefani@df.uba.ar (F.D.S.), mauricio.pilopais@unifr.ch (M.P.P.), guillermo.acuna@unifr.ch (G.P.A.).

**Optical nanoantennas are structures designed to manipulate light-matter interactions at the nanoscale by interfacing propagating light with localized optical fields. In recent years, a plethora of devices have been realized that are able to efficiently tailor the absorption and/or emission rates of fluorophores. By contrast, modifying the spatial characteristics of their radiation fields remains a challenge. Up to date, the designs providing directionality to fluorescence emission have required compound, complex geometries with overall dimensions comparable to the operating wavelength. Here, we present the fabrication and characterization of DNA-templated ultracompact optical antennas, with sub-wavelength sizes and capable of directing single-molecule fluorescence into predefined directions. Using the DNA origami methodology, two gold nanorods are assembled side-to-side with a separation gap of 5 nm. We show that a single fluorescent molecule placed at the tip of one of the nanorods drives the dimer antenna in anti-phase, leading to unidirectional emission.**

Light-matter interactions at the nanoscale are limited by the two orders of magnitude size mismatch between the electronic confinement of quantum emitters (in the nm range) and the wavelength of the optical radiation (in the hundreds of nm range). In analogy to their microwave or radiofrequency counterparts, optical antennas can act as transducers between propagating light and localized fields. To date, a plethora of optical antennas have been designed and demonstrated[1,2], improving by several orders of magnitude the rates of light absorption and emission of individual photon emitters[3]. Specifically, plasmon-induced localized fields have been extensively exploited to enhance single-molecule spectroscopic signals including fluorescence[4] and Raman scattering[5]. A less studied aspect of optical antennas is their ability to impose directionality of the emitted light[6].

Photon absorption and emission involve dipolar, or under special conditions multipolar[7], transition moments, and therefore do not possess intrinsic unidirectionality. Thus, artificial schemes to increase the emission directionality of photon emitters are of crucial significance, for example in the context of integrated optical circuits and quantum communication/computing schemes[8,9].

Over a decade ago, Curto *et al.* used lithographically-made Yagi-Uda antennas to demonstrate the directional emission of quantum dots[10]. A drawback of such phased array antennas is that they require placing various elements at distances of around $\lambda/4$[11,12], which extends their footprint to dimensions larger than the wavelength. Several more compact directional designs aiming to reduce the overall effective size by an order of magnitude have been proposed and numerically investigated[13]. A scheme used by such ultracompact directional antennas exploits the interplay of spectrally overlapping electric and magnetic resonances of single high-index dielectric nanoparticles[14], or arrangements of metallic nanoparticles that can support both electric and magnetic modes in the same spectral range[15]. Alternatively, Pakizeh *et al.*[16] proposed a design of two gold nanodisks separated by 10 nm. A photon emitter in the vicinity of one of the disks drives the resonance of the two metallic disks with a phase delay between each other introduced by the separation gap, resulting on unidirectional emission at frequencies close to the anti-phase plasmon mode of the system[16].

Even though the potential and working principles of directional ultracompact antennas (DUAs) for single emitters have been long known, their experimental realization remained elusive due to several shortcomings of traditional top-down nanofabrication methods. For instance, top-down methods can hardly produce the reduced gaps under 20 nm necessary for ultracompact antennas, and more importantly, they do not allow placing single emitters near an optical antenna element with positional and stoichiometric control.

In this work, we report the realization of an ultracompact directional antenna for single-photon emitters produced by means of DNA self-assembly. The DNA origami technique[17–19] was used to assemble two colloidal gold nanorods (AuNRs) in a side-to-side arrangement with a controlled separation gap, and a single fluorescent molecule at the tip of one of the AuNRs.

The geometry and working principle of a DUA driven in anti-phase is schematically shown in Figure 1A, including the two AuNRs and the single dipolar emitter (black arrow) placed at the tip of one of the AuNR elements. The overall emission can be approximated by the sum of the induced dipolar oscillations in each element (blue and red arrows). Depending on the geometry, materials, and wavelength (frequency of the emitter), conditions may be met where constructive and destructive interference occur at opposite sides of the DUA, leading to unidirectional emission. To illustrate this, Figure 1B shows schematically the calculated angular emission pattern of a DUA driven in anti-phase along with the emission pattern of a single AuNR (monomer antenna).

To fabricate such DUAs, we used a T-shaped DNA origami host structure (Figure 1C) as a template to assemble previously functionalized colloidal AuNRs, through DNA hybridization, in a side-to-side configuration (Figure 1D) with a separation gap of 5 nm (further details are included in Figure S1 and Table S1). The origami design enables the incorporation of a single fluorescent molecule at the tip of one of the AuNRs, 5 nm away from the gold surface (red dot in Figures 1C and D). Commercial AuNRs with nominal dimensions of 40 nm in diameter and 68 nm in length were self-assembled in solution with the DNA origami to form the DUAs. The DUAs were then purified by gel electrophoresis and imaged by transmission electron microscopy (TEM), confirming the correct self-assembly of our structures (Figure 1E). Some degree of variability in the dimensions of the AuNRs is observed, in agreement with the size distribution of the colloid.

Considering these dimensions, DUAs have an overall size of 85×73 nm$^2$, which corresponds to an order of magnitude smaller than phased-array (Yagi-Uda) configurations[11,12].

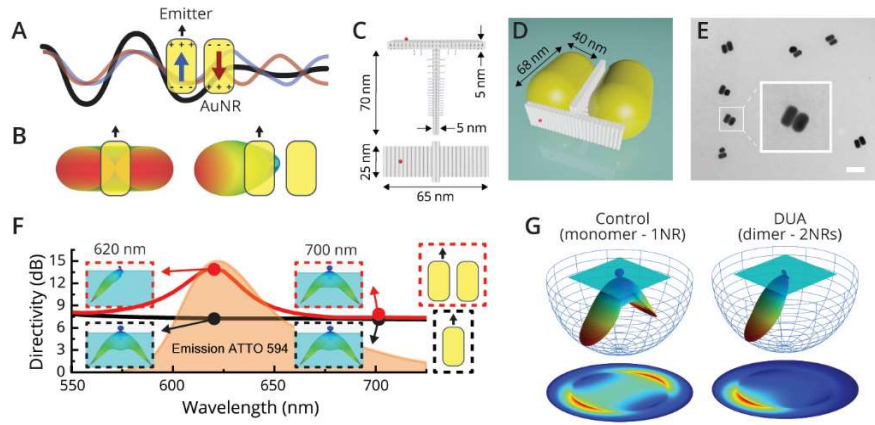

*Figure 1. Geometry and working principle of a directional ultracompact antenna (DUA) based on two gold nanorods (AuNRs) placed side-to-side. (A) If the quadrupole-like anti-phase mode is excited by a dipolar source (black arrow) in the near-field of one element, constructive and destructive interferences can be achieved at opposite sides of the antenna, leading to unidirectional emission. (B) Calculated angular emission patterns for a single AuNR (monomer antenna) and a DUA (dimer) driven by a single dipolar source in vacuum. (C, D) Sketch of the DNA origami host structure (C) and the assembled antenna (D) with a single photon emitter (red dot). (E) TEM images of DUAs, scale bar 200 nm. (F) Spectral directivity for both a DUA with the geometry described in (C) and a monomer antenna placed on a glass substrate, together with the emission spectra of ATTO 594. The insets show the emission pattern close to the anti-phase mode wavelength (620 nm) and away from it (700 nm). (G) Simulated back-focal plane images for the monomer antenna and the DUA.*

The spectral response of the fabricated DUAs on a glass substrate was investigated numerically by means of finite element simulations. The directivity vs. wavelength and exemplary emission patterns of a DUA (red line) and a monomer antenna (black line) are shown in Figure 1F. For an ideal DUA composed of two identical AuNRs, strong unidirectional emission is predicted for emitters with an emission wavelength from 600 to 630 nm, where the DUA is driven in anti-phase[20]. These wavelengths are red-shifted from both the longitudinal resonance of a single AuNR and the in-phase resonance of the DUA (see Figure S2). Thus, a fluorophore emitting in this range (e.g ATTO 594, Figure 1F), is expected to show unidirectional emission when coupled to the DUA, for instance detectable by the very distinct back-focal plane (BFP) images[21] when compared to the dipolar emission from the monomer antenna control (Figure 1G).

The performance of the fabricated DUAs was studied experimentally in a custom-built scanning fluorescence microscope modified to relay the BFP image of the structure into a CCD (see details in Methods). For each DUA, a series of BFP images was acquired, from which an intensity transient was generated (Figure 2A). Only transients with a single bleaching step, a signature of emission arising from a single fluorophore, were employed for further analysis. This approach, in addition to confirming that only one molecule was driving the antenna, enabled the discrimination of the single-molecule fluorescence (FL) emission from photoluminescence (PL) of AuNRs[22]. This analysis was combined with scanning electron microscopy (SEM) imaging to correlate the antenna structure with the directionality of single-molecule emission.

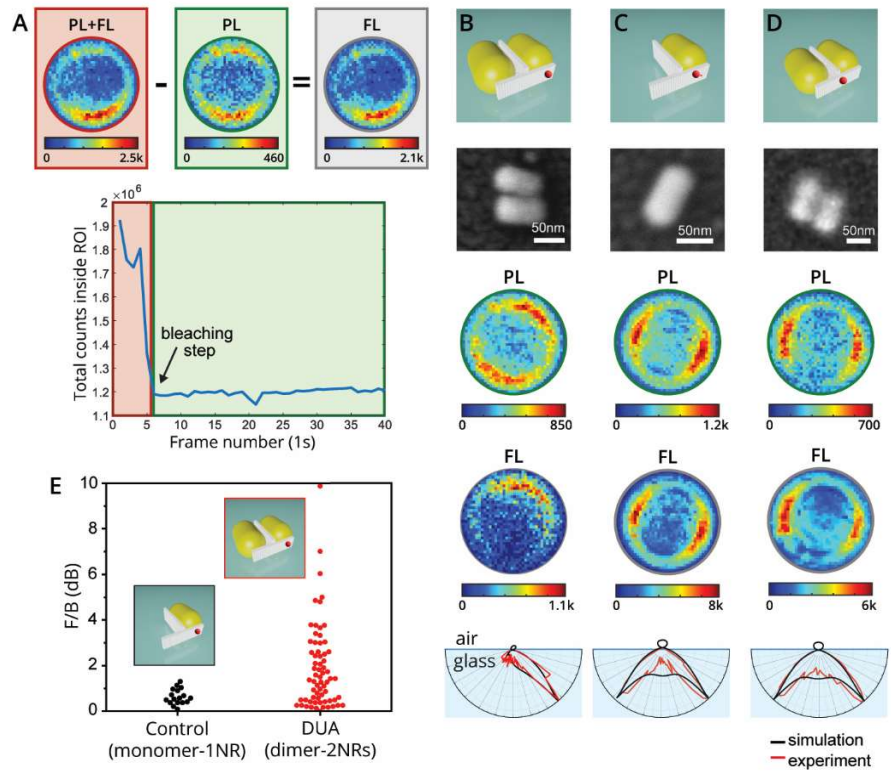

*Figure 2. **Analysis pipeline and antennas´ performance**. (A) Determination of fluorescence (FL) and photoluminescence (PL) patterns from back-focal plane (BFP) imaging. For each antenna, a series of BFP images were acquired. From the integrated intensity transient, the time point where the single ATTO 594 fluorophore bleaches is identified. The BFP image corresponding to FL emission by the single fluorophore is obtained by subtracting the averaged BFP image before (PL+FL) and after (PL) bleaching. (B-D) Performance examples of a directional ultracompact antenna (DUA) (B), a monomer antenna (C), and a dimer antenna with the fluorophore in the center (D). From top to bottom, 3D sketch of the structure, SEM image, BFP PL and FL images, and emission polar patterns. The position of the single fluorophore is indicated by a red dot. (E) Bee swarm distribution plots of the forward-to-backward (F/B) ratios obtained from the BFP patterns of both a monomer and a DUA.*

An exemplary result for a DUA is shown in Figure 2B. The PL BFP image shows a symmetric pattern with two lobes of similar intensity, separated by a line of minimum intensity parallel to the AuNRs, as determined from the SEM image. This emission pattern is characteristic of a dipolar emitter and can be rationalized based on the symmetry of the structure and the fact that PL emission corresponds to far-field excitation of the in-phase plasmon dipolar mode. The situation changes for the FL emission when the DUA is driven in anti-phase by the single fluorophore in the near field. The FL BFP image shows a single lobe of high intensity indicating unidirectional emission. More examples of DUAs are shown in Figure S3.

The directionality achieved with this ultracompact design depends critically not only on the positioning of the antenna elements but also on the position of the emitter. To demonstrate this, we exploited the unique flexibility and nanometer positioning control of the DNA origami technique to self-assemble a monomer antenna (Figure 2C) and a symmetric dimer antenna where the single ATTO 594 fluorophore is laterally shifted to an equidistant position from the two AuNRs (Figure 2D). In these two cases, both BFP images of PL and FL are symmetric with two main lobes as expected for a dipolar emitter. The experimentally determined angular emission for the three antenna designs are in excellent agreement with the calculations (lower panels of Figures 2B-D).

Another advantage of nanofabrication based on DNA self-assembly is that it enables the parallel production of zillions of nominally identical structures. We characterized 67 DUAs and 17 monomer antennas, and quantified their performance in terms of the forward-to-backward ratio (F/B) from the BFP images using the so-called areal method[23,24] (see further details on Methods). Figure 2E shows the fluorescence F/B distributions in dB. Ideally, the monomer antennas should have F/B = 0 dB. In reality, they present values between 0 and 1.3 dB. We ascribe this residual asymmetric emission to measurements with limited signal-to-noise ratio, and slightly irregular geometry of AuNRs. In contrast, the DUAs present a much broader distribution of F/B values, spanning a range from 0 to almost 10 dB. The maximum directionally obtained, 9.9 dB, is higher than the best reported values for top-down lithographic Yagi-Uda antennas, either driven by a single quantum dot (6 dB)[10] or electrically (9.1 dB)[23]. Several reasons explain the observed distribution of F/B[25]. In addition to the factors mentioned for the monomer antennas, the DUAs performance may also be affected by the size distribution of the colloidal AuNRs, because variations in the size of the AuNRs modifies both the magnitude of the directivity as well as the spectral operating range (more in Figure S4).

In summary, we have experimentally demonstrated that the emission of single-photon sources can be mediated with high degree of unidirectionality by means of an ultracompact optical antenna based on two metallic nanorods in a side-to-side configuration with deep subwavelength separation gaps (~$\lambda$/50). This approach, theoretically predicted over a decade ago, is not based on an antenna phased array or the interplay between magnetic and electric resonances. Instead, it uses the near-field excitation of the anti-phase mode by a dipolar emitter precisely located on top of one of the antenna elements. This type of antennas shows a critical spectral dependency of the directionality on the geometry of the antenna. We circumvented these challenges by an alternative, bottom-up parallel self-assembly technique using a T-shaped DNA origami structure to host two colloidally prepared AuNRs, together with a single organic fluorophore. Our results confirm the theoretical predictions and show that ultracompact unidirectional optical antennas can be realized when the nanofabrication is sufficiently fine controlled, in terms of antenna elements geometry and position of the photon emitter. While unidirectionality was obtained when the fluorophore was placed on top of one the AuNRs, it was lost when the fluorophore was shifted sideways just 22.5 nm, to an equidistant position from the two AuNRs, or if only one AuNR was present. Furthermore, we achieved a maximum unidirectionality of almost 10 dB, outperforming previously reported top-down lithographic Yagi-Uda antennas based on five elements and with a footprint an order of magnitude larger.

We propose that this work, in combination with recent advances and current efforts in DNA self-assembly[26,27], will open new routes for the fabrication of more complex nanophotonic devices, circuits, and light-emitting metasurfaces[28] at large scales[29,30].

**Methods**

**Numerical simulations**

Three-dimensional full-wave simulations were carried out using frequency domain solver of CST Studio Suite.

AuNRs were modeled as cylinders (R=20 nm, L=52 nm) with two semi-ellipsoidal caps (a=b=R, c=8 nm), making the total length 68 nm. For the dielectric function of gold, we used fitting data from Johnson & Christy. AuNRs were uniformly coated with 2 nm DNA layer ($n_{ssDNA}$ = 1.7). The gap between the two coated AuNR surfaces was set as 5 nm. The distance between substrate and coating layer was set as 5 nm. The surrounding medium was set to air ($\varepsilon$ = 1).

The scattering spectra of AuNRs on a glass substrate ($\varepsilon$ = 2.25) were obtained by subtracting the scattered field in the absence of AuNRs from the scattered field in the presence of the AuNRs. Plane waves with TE and TM polarization were used as the excitation sources at normal incidence.

The final simulated scattering spectra was then obtained by averaging according to $C_{sca} = \frac{1}{2}(C_{sca}^{TE} + C_{sca}^{TM})$.

For the far field pattern, a Hertzian dipole (oriented along the long axis of AuNR) was simulated as one electric dipole emitter. This dipole was placed 3 nm away from the DNA coating layer of AuNR. The directivity of an antenna is defined as $Dir_{max} = \frac{4\pi S_{max}(\theta, \varphi)}{\int_0^{2\pi} \int_0^{\pi} S(\theta, \varphi) \sin\theta d\theta d\varphi}$, where $S(\theta, \varphi)$ represents the power radiated by the antenna in a given direction $(\theta, \varphi)$ per unit solid angle. In order to facilitate comparison with experimental BFP images, far-field radiation into the lower half space was projected into the Fourier plane to get 2D patterns.

**Forward to Backward ratio quantification**

Forward to Backward ratio (F/B), defined as the integral ratio of radiated power $S(\theta, \varphi)$ in two angular ranges (($\theta_1 - \delta_1 \rightarrow \theta_1 + \delta_1$, $\varphi_1 - \delta_2 \rightarrow \varphi_1 + \delta_2$) and ($\theta_2 - \delta_1 \rightarrow \theta_2 + \delta_1$, $\varphi_2 - \delta_2 \rightarrow \varphi_2 + \delta_2$)), is used to quantify unidirectionality of antenna:

$$F/B = 10\log_{10} \frac{\int_{\theta_1-\delta_1}^{\theta_1+\delta_1} \int_{\varphi_1-\delta_2}^{\varphi_1+\delta_2} S(\theta, \varphi) \sin\theta d\theta d\varphi}{\int_{\theta_2-\delta_1}^{\theta_2+\delta_1} \int_{\varphi_2-\delta_2}^{\varphi_2+\delta_2} S(\theta, \varphi) \sin\theta d\theta d\varphi} \quad (dB)$$

Here, $(\theta_1, \varphi_1)$ corresponds to the angular position of the region of maximum signal (forward direction) whereas $(\theta_2, \varphi_2)$ is the position of the opposite sector (backward direction), with $\theta_2 = \theta_1$ and $\varphi_2 = \varphi_1 + \pi$. Given the signal distribution in the observed BFP patterns, we chose $2\delta_1$ to cover a 6 super-pixels distance, $\delta_2=25°$ (total angular range of 50°). Given the F/B ratio in the simulated BFP patterns, we chose $\delta_1=10°$.

**Single molecule measurements**

A sample-scanning confocal microscope with a high NA objective (Olympus, 100× NA=1.4) was modified to include a flip mirror in the detection path to switch between confocal and BFP imaging. Briefly, the BFP imaging path consists of a relay lens (f=100 mm) and a Bertrand lens (f=50 mm)[31], that produce a BFP image onto a CCD sensor (Andor iXon Ultra 888). The confocal detection is done with a single-photon counting avalanche photodiode (τ-SPAD, PicoQuant).

Measurements started with a confocal image of a region of interest in the sample from which the location of the target structures is determined. Then, the detection is switched to BFP imaging and the sample is moved to position a single structure in the center of the excitation focus. The CCD was used in kinetic mode in order to acquire videos. An integration time of 1s per frame was applied, and a sufficient number of frames was recorded (~60 frames) to allow for the observation of the bleaching step and to record PL BFP images after that. A 4×4 binning was applied during acquisition to have a higher signal per pixel[21]. This results in an emission pattern consisting of 36×36 super-pixels (1.9×1.9 mm). To achieve a good SNR, the EM gain of the camera was set to 75.

For analysis of the BFP videos, the total signal inside a circular ROI containing the pattern is retrieved and plotted as a function of the frame number. This yields an intensity time trace (see Figure 2A) that can be used to identify the bleaching frame. Frames are averaged before (PL+FL) and after (PL) this bleaching step, and the subtraction of these two averaged images yields the average single-molecule signal (FL).

**DNA origami design and folding**

The T-shape DNA origami was designed using CaDNAno[32] and visualized for twist correction using CanDo[33]. Figure S1 shows a cartoon of the origami indicating its dimensions and the positions of the modified staples. In short, the DNA template has 18x Poly-A8 (blue) and 2x Poly-A12 (green) ssDNA strands (Biomers.net GmbH) at each side that serve as handles to accommodate nanoparticles surface functionalized with Poly-T. An ATTO 594 fluorophore (red, Biomers.net GmbH) is positioned directly on top of one of the NRs, or at the top-center, as shown on Figure S1. The design files are uploaded at https://nanobase.org/[34].

Unmodified DNA sequences were purchased from Integrated DNA Technologies, INC. In short, a scaffold consisting of a vector derived from the single-stranded M13-bacteriophage genome (M13mp18, 7249 bases, Bayou Biolabs) and staples (100 nM, ca. 32 nts) were mixed in a 1x TAE buffer (40 mM Tris, 10 mM Acetate, 1 mM EDTA, pH 8, stock purchased from Alfa, CAS#77-86-1, J63931.k3) containing 12 mM $MgCl_2$ (stock purchased from Alfa, CAS#7786-30-3, J61014.AK). The solution was heated to 75°C and ramped down to 25°C at a rate of 1 degree every 20 mins. The folded DNA origami structures were purified from excess staple strands by gel electrophoresis using a 0.8 % agarose gel (LE Agarose, Biozym Scientific GmbH) in a 1x TAE buffer/12 mM $MgCl_2$ buffer for 2.5 hours at 4 V/cm. The appropriate band containing the targeted DNA template was cut out and squeezed using cover slips wrapped in parafilm.

**Nanorods functionalization and attachment to DNA template**

Experiments were performed with AuNRs prepared following previously established protocols[35] (further details are available in the Supporting Information) and with commercial nanorods with nominal dimensions of 40 nm diameter and 68 nm length (Nanopartz inc.). Functionalization of both custom built and commercial AuNRs was performed in the same way. Thiolated DNA (Thiol-C6-T18, ELLA Biotech GmbH) was mixed with Au nanorods at a 100 nmol:100 uL_100OD ratio and frozen overnight[36]. Excess DNA was removed using gel electrophoresis. This step also ensures the removal of any self-aggregated dimer formed during the NP functionalization. The concentration of AuNRs was determined via UV-Vis absorption spectroscopy (Nanodrop).

The purified DNA template was mixed with the purified AuNRs using an excess of five AuNRs per binding site and adding NaCl to a final concentration of 600 mM. The sample was then annealed at 30°C and cooled down to 25°C at a rate of 1 degree every 30 min. This cycle was repeated 6 times. The excess of NPs was then removed by gel electrophoresis (running for 4.5 hours) and the band containing correctly formed dimers was extracted as described before.

**Sample preparation**

For immobilization of the structures, glass cover slips were first rinsed with water and then cleaned in a UV cleaning system (PSD Pro System, Novascan Technologies, USA). The surfaces were functionalized by sequentially immersing the substrates in the following solutions: BSA-biotin/PBS (0.5 mg/mL, Sigma Aldrich, CAS#9048-46-8, A8549-10MG) for 20 mins, neutrAvidin/PBS (0.5 mg/mL, Thermofischer, 10443985) for 20 min, and 5' biotin-Poly-A15-3' ssDNA for 20 mins (Biomers.net GmbH). In between all the steps, the coverslips were washed with 1x PBS buffer (Alfa, J75889.K2). The so prepared substrates were incubated overnight with the purified structures of DNA-origami with AuNRs for their immobilization through DNA hybridization of the Poly-A on the surface and the free Poly-T on the AuNRs.

**Author contributions**

**Acknowledgements**

L.M.L.-M. acknowledges funding from the European Union's Horizon 2020 research and innovation program under grant agreement No 861950, project POSEIDON and MCIN/AEI /10.13039/501100011033 through Maria de Maeztu Unit of Excellence No. MDM-2017-0720. X. Z. acknowledges funding from the Juan de la Cierva Fellowship (FJC2018-036104-I). F.D.S. acknowledges the support of the Max Planck Society and the Alexander von Humboldt Foundation. This work has been funded by Consejo Nacional de Investigaciones Científicas y Técnicas (CONICET) and Agencia Nacional de Promoción Científica y Tecnológica (ANPCYT), projects PICT-2017-0870, and PICT-2014-3729. A.F.D. acknowledges funding from the Spanish MCIN/AEI/10.13039/50110001033 and by "ERDF A way of making Europe" through Grant No. RTI2018-099737-B-I00, and the 2020 CAM Synergy Project Y2020/TCS-6545 (NanoQuCo-CM). G.P.A. acknowledges support from the Swiss National Science Foundation (200021_184687) and through the National Center of Competence in Research Bio-Inspired Materials (NCCR, 51NF40_182881).

# Supporting Information

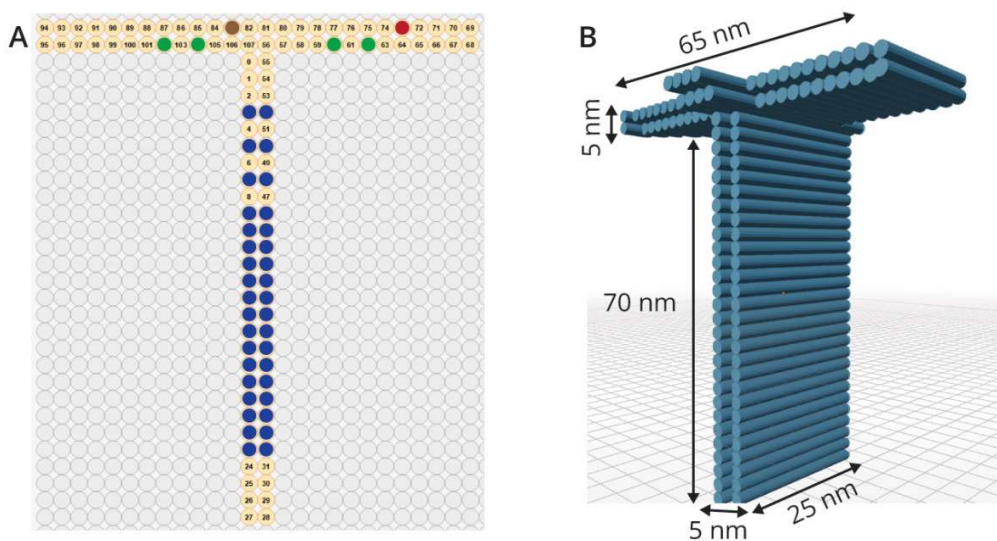

***Figure S1: T-shape DNA origami design.*** *(A) Cartoon representation of the T-shape DNA template. Poly-A handles (A8 in blue and A12 in green) serve to assemble two 40×68 nm AuNRs in a parallel configuration, while spaced by 2 helix-bundles (5 nm). The DNA-template can accommodate a single ATTO594 fluorophore either on top of one of the nanorods (red) or between the two rods (brown) to form the different configurations as explained in the main text. (B) 3D view of the DNA origami, including dimensions.*

| | |
|---|---|
| Atto594 Position Center | 5' aagaataggaacgtggggcacagacaatatttt **[Atto594]** 3' |
| Atto594 Position Right | 5' catgtcaataagcaaaataatcctgattgttttt **[Atto594]** 3' |

***Table S1:*** *List of modified staples for the respective DNA-origami designs.*

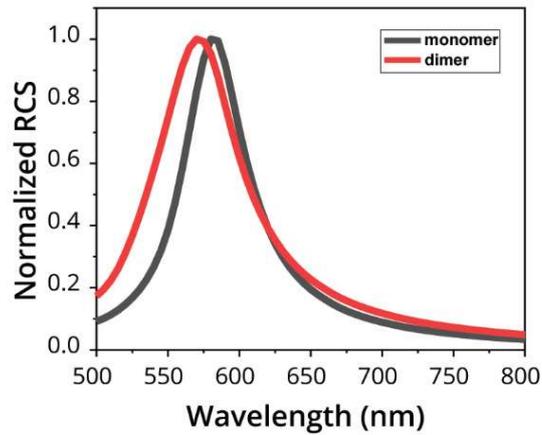

*Figure S2: Simulated scattering spectra of monomer and dimer antennas. The size of the AuNRs is set to 40×68 nm, and the gap between them is 5 nm. The in-phase mode observed in the dimer spectrum is blue-shifted with respect to the monomer resonance. The spectral window where we expect to have maximum F/B is red-shifted from both the longitudinal resonance of monomer antennas and the in-phase mode of the dimer shown here, confirming that this unidirectional behavior is coming from the excitation of the anti-phase mode by the emitter.*

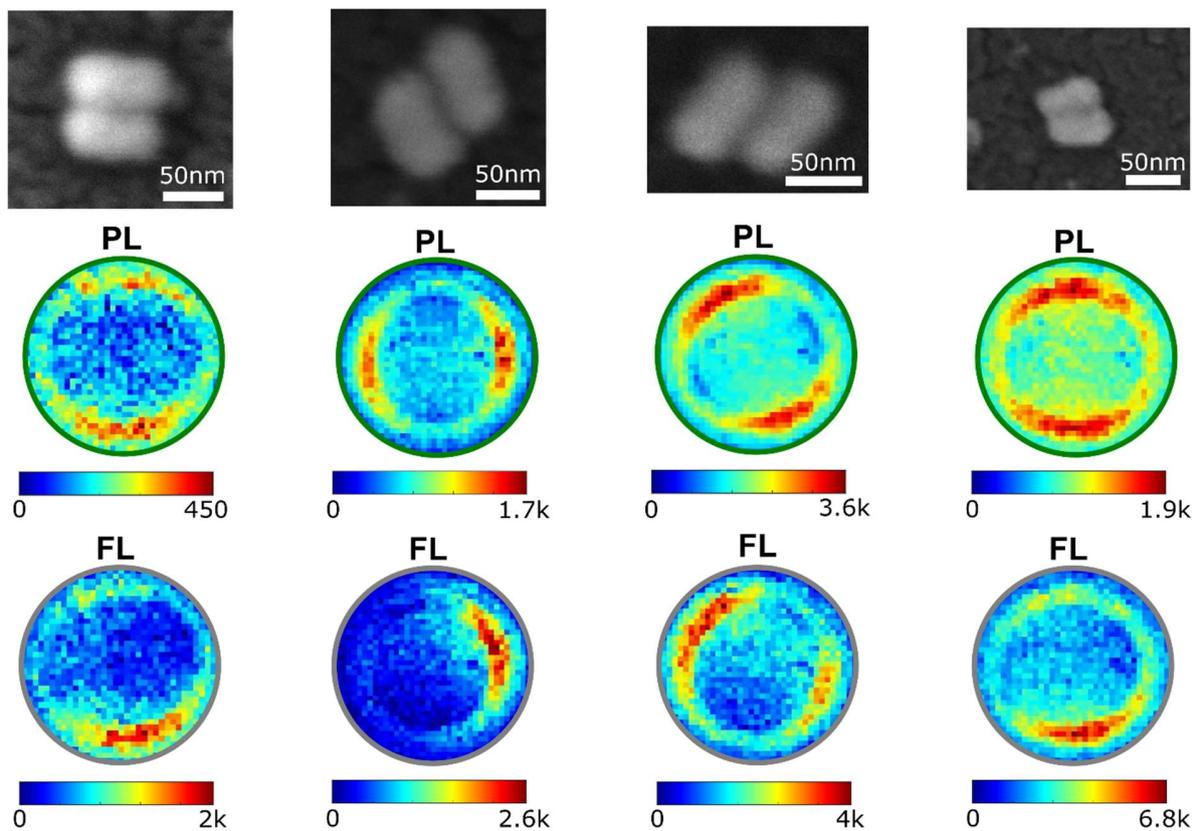

*Figure S3: Examples of directional ultra-compact antennas.* Representative SEM and BFP images of four different ultra-compact antennas, showing good reproducibility in the assembly and unidirectional single-molecule emission.

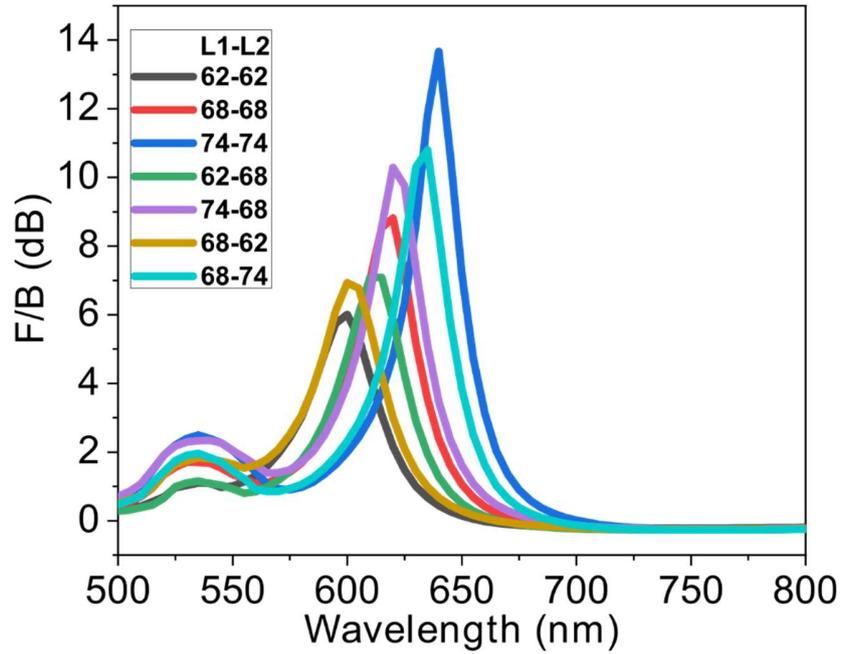

*Figure S4: Simulation of F/B ratio vs. wavelength for directional ultracompact antennas made with AuNRs of different dimensions. BFP patterns of ultracompact directional antennas were simulated for different lengths of the AuNRs (diameter was fixed at 40 nm). The F/B was extracted using the same angular range used for the experimental data. L1 corresponds to the length of the AuNR closer to the emitter, and L2 is the length of the second nanorod.*

**Synthesis of AuNRs**

For the synthesis of AuNRs with a size of 50×82 nm, we applied a modified seeded growth method using cetyltrimethylammonium bromide (CTAB) and n-decanol as binary surfactants[1]. Growth solutions A (CTAB 50 mM, n-decanol 13.5 mM) and B (CTAB 50 mM, n-decanol 11 mM) were prepared by dissolving 9.111 g of CTAB together with 1.068 g and 870.5 mg of n-decanol in 500 mL of warm water, respectively. Both growth solutions were stirred at 50 °C for at least 1 h and then cooled down to 25 °C in a water bath. The synthesis involves three steps. First, a freshly prepared ascorbic acid solution (0.1 M, 25 μL) and a freshly prepared NaBH4 solution (0.02 M, 0.2 mL) were successively added to a mixture of growth solution A (5 mL) and HAuCl4 (0.05 M, 0.05 mL). The reaction solution turned from light yellow to brownish, indicating the formation of 1−2 nm Au seeds. The seed solution was aged in a water bath at 25 °C for 30 min under mild stirring. Second, the aged seed solution (0.6 mL) was injected into an aqueous growth solution composed of growth solution A (10 mL), HAuCl4 (0.05 M, 100 μL), AgNO3 (0.01 M, 80 μL), HCl (1 M, 700 μL) and ascorbic acid (0.1 M, 130 μL) under vigorous stirring. The reaction solution was kept in a water bath at 25 °C for at least 4 h under stirring, during which small Au NRs of 20 nm × 7 nm were formed. The as-prepared small Au NRs were washed by CTAB solution (10 mM) and centrifugation (14000 rpm, 45 min) for twice, and finally concentrated to OD400 = 10. Third, the small Au NR solution (130 μL) was injected into an aqueous growth solution composed of growth solution B (80 mL), HAuCl4 (0.05 M, 800 μL), AgNO3 (0.01 M, 1.6 mL), HCl (1 M, 400 μL) and ascorbic acid (0.1 M, 640 μL) under vigorous stirring. The reaction solution was kept in a water bath at 28 °C for more than 4 h under stirring. As-prepared Au NRs were centrifuged at 3000 rpm for 20 min twice and redispersed in CTAB (1 mM).

Chemicals. Tetrachloroauric acid (HAuCl4·3H2O, ≥99%), citric acid (≥ 99.5%), sodium borohydride (NaBH4, 99%), L-ascorbic acid (AA, ≥99%), silver nitrate (AgNO3, ≥99%), 1-decanol (n-decanol, 98%), and hexadecyltrimethylammonium bromide (CTAB, ≥99%) were purchased from Sigma-Aldrich. Hydrochloric acid solution (HCl, 37 wt%) was purchased from Scharlau. All chemicals were used without further purification. Milli-Q water (resistivity 18.2 MΩ·cm at 25 °C) was used in all experiments. All glassware was cleaned with aqua regia, rinsed with Milli-Q water, and dried before use.